\documentclass[pre,aps,twocolumn,superscriptaddress,amsmath,showpacs,floatfix]{revtex4-1}
\usepackage{graphicx}

\begin{document}
 \title{First Passage Distributions in a Collective Model of
 Anomalous Diffusion with Tunable Exponent}
 \author{Assaf Amitai}
 \affiliation{Raymond and Beverly Sackler School for Physics and
 Astronomy,  Tel Aviv University, Tel Aviv 69978, Israel}
 \author{Yacov Kantor}
 \affiliation{Raymond and Beverly Sackler School for Physics and
 Astronomy,  Tel Aviv University, Tel Aviv 69978, Israel}
 \author{Mehran Kardar}
 \affiliation{Department of Physics, Massachusetts Institute of
 Technology, Cambridge, Massachusetts 02139, USA}

 \date{\today}

\begin{abstract}
We consider a model system in which anomalous diffusion is generated by
superposition of underlying linear modes with a broad range of relaxation
times. In the language of Gaussian polymers, our model corresponds to
Rouse (Fourier) modes whose friction coefficients scale as wavenumber to
the power $2-z$. A single (tagged) monomer then executes subdiffusion over
a broad range of time scales, and its mean square displacement increases
as $t^\alpha$ with $\alpha=1/z$. To demonstrate non-trivial aspects of the
model, we numerically study the absorption of the tagged particle in one
dimension near an absorbing boundary or in the interval between two
such boundaries. We obtain absorption probability densities as a
function of time, as well as the position-dependent  distribution for
unabsorbed particles, at several values of $\alpha$. Each of these
properties has features characterized by exponents that depend on $\alpha$.
Characteristic distributions found for different values of $\alpha$ have
similar qualitative features, but are not simply related quantitatively.
Comparison of the motion of translocation coordinate of a polymer moving
through a pore in a membrane with the diffusing tagged monomer with identical
$\alpha$ also reveals quantitative differences.

\end{abstract}

\maketitle

\section{Introduction}

In many physical processes one encounters a stochastic variable whose mean
squared fluctuations increase with time $t$ as $t^\alpha$ with $\alpha\ne 1$.
These processes are sometimes referred to as anomalous diffusion~\cite{fbmgle},
and specifically {\em subdiffusion} for $\alpha<1$. Such behavior is usually
caused by the collective dynamics  of numerous degrees of freedom,
or modes with a broad distribution of characteristic times. The exact
relations between the underlying modes and the observed coordinate are
usually unknown, and first-principle derivation of the equations governing the
anomalous diffuser are rare.  As a result, a variety of such processes are
typically grouped into broad classes in accordance
to their general characteristics~\cite{mkreview}. While the exponent $\alpha$
is an important and convenient indicator of anomalous dynamics, it
contains little information on the hidden underlying driving forces.

A different set of properties of a diffuser can be revealed by observing
its {\em first passage} to a target, or its absorption at a
trap~\cite{feller,redner,weiss}.  For instance, it has been
established~\cite{fpinf} that the probability density function (PDF) $Q(t)$
to be absorbed for  one-dimensional (1D) subdiffusion between two absorbing
walls has a power law tail, if the process is described by a fractional
diffusion equation~\cite{mkreview}.  This slow decay of $Q(t)$ for a
subdiffuser following such dynamics leads to a diverging mean absorption
time. On the other hand, Kantor and Kardar~\cite{kkgauss} demonstrated that
a monomer in a Gaussian polymer that is characterized by $\alpha={1}/{2}$ has
a finite absorption time when it diffuses between two absorbing boundaries.
The presence of absorbing boundaries introduces additional
characteristics, such as the long time behavior of the survival
probability $S(t)$ in the presence of a single absorbing wall, or
the behavior of the PDF of particle position near the trap.
In specific cases with a well-defined $\alpha$ (see, e.g., Refs.~\cite{zk,zoia})
some aspects of absorption characteristics have been determined.
However, in the absence of rigorous scaling relations one cannot establish
whether the exponent $\alpha$ determines all other characteristics.
Recently, Zoia {\em et al.}~\cite{zrm} used some general scaling arguments
to propose such a relation, furthering the need to probe such properties
with tunable exponent $\alpha$.

While it is generally recognized that underlying (hidden) processes are responsible
for anomalous dynamics, many theoretical approaches simplify the problem to
effective equations for the observed variable, hoping to capture the multitude of
underlying interactions. There is no {\em a priori} reason for such an approach
to succeed, and it is therefore useful to consider alternative models where the
underlying processes are well characterized.
In this work we consider a model in which subdiffusion is generated as a result
of superposition of underlying modes with a broad range of time scales.
For solvability and ease of simulation we limit ourselves to linear (but stochastic)
dynamics for the modes.
Our model is closely related to the dynamics of a Gaussian polymer, or to
fluctuations of a Gaussian interface.
In the polymer language, which we adapt for most of the presentation,
the anomalous behavior of a  single (tagged) monomer \cite{KBG} is easily understood
in terms of the superposition of underlying Rouse modes.
The resulting exponent $\alpha$ depends on whether the polymer dynamics
is diffusive (Rouse) or influenced by hydrodynamic interactions (Zimm).
The difference between the two cases can be cast as due to a wave-length
dependent friction coefficients.

In Sec.~\ref{sec:mod} we take this analogy a step further, and show that any
value of $\alpha$ can be generated by appropriate scaling of the friction
coefficients with wavelength.
While we adapt notions from polymer physics in developing our model, the approach we take is not intended to address any particular polymer problem. Rather, we rely on this well-defined mathematical model to explore issues pertinent to anomalous diffusion (specifically absorption), and to compare and contrast with other mathematical models introduced in this context.
Our approach is closely related to the model
proposed by Krug {\em et al.} \cite{krug} where the value of $\alpha$
is controlled by modifying the forces between particles.
We use our generalization to study anomalous diffusion in the presence
of one and two absorbing boundaries in Secs.~\ref{sec:abs}
and \ref{sec:pdfs}. In particular, we explore the
long-time tails of the absorbtion probability $Q(t)$, as well
as the asymptotic stable shapes of the PDFs of the surviving walkers.
Qualitatively, various quantities have similar features for a variety of
$\alpha$. However, we find that $\alpha$ does not enter the results in a
trivial way, and the stable function for one $\alpha$ cannot be
obtained from another by simple transformation. Moreover,
the comparison of results for our walker with exponent
$\alpha$ coinciding with that of a polymer translocating
though a membrane pore (Sec.~\ref{sec:transloc}) demonstrates quantitative
differences between the two cases. Some possible extensions of this work
are discussed in Sec.~\ref{sec:disc}.

\section{Model}
\subsection{Rouse modes of polymers and anomalous dynamics of a monomer}\label{sec:pol}

Polymers~\cite{deGennes_book} provide a relatively simple physical system
in which the collective motion of monomers leads to behavior spanning a
broad range of time scales~\cite{doi}.
Ignoring the interactions between non-covalently-bonded monomers, the dynamics
of a polymer can be reduced to independent Rouse (Fourier) modes~\cite{rouse}.
For a polymer consisting of $N$ monomers, each such mode $U_p$
($p=0,\cdots,N-1$) has a distinct relaxation time $\tau_p$.
The Gaussian (or ideal) polymer is a particularly simplified model  composed of
beads (monomers) connected by harmonic (Gaussian) springs.
Each polymer configuration is now described by the set of monomer positions $R_n$
($n=1,2,\cdots,N$), and has energy (again neglecting further neighbor interactions)
\begin{equation}\label{eq:Hamiltonian}
H = \frac{\kappa}{2} \sum_{n=1}^N \left( R_n - R_{n-1} \right)^2\,.
\end{equation}
Here the spring constant is $\kappa = k_B T/b^2$, where $k_B$ is
the Boltzmann constant, $T$ is the temperature, and $b$
is the root-mean-square distance between a pair of connected monomers.

From Eq.~\eqref{eq:Hamiltonian} we can construct a simple relaxational
(Langevin) dynamics for the Gaussian polymer, whereby
\begin{equation}\label{eq:Rouse_Langevin}
\zeta \frac{d {R}_n}{dt} = -\kappa(2{R}_n-{R}_{n+1}-{R}_{n-1}) +{f}_n\,,
\end{equation}
for $1<n<N$. The deterministic force (first term on the right hand side) is different for
$R_1$ and $R_N$, since the end monomers are attached only to a single neighbor.
Here, $\zeta$ is the friction coefficient of the monomer, while the noise $f_n(t)$
has a zero mean and correlations of
$\langle {f}_{n}(t){f}_{m}(t') \rangle = 2\zeta k_{B}T\delta_{n,m}\delta(t-t{'})$,
to ensure proper thermal equilibrium at temperature $T$.
In one dimension the positions $\{R_n\}$ are scalars, while the generalization to
vectorial coordinates in higher dimensions is trivial, as in this model the coordinates
in different spatial dimensions are  independent.

Rouse modes are now obtained by Fourier transformation as~\cite{doi}
\begin{equation}\label{eq:def_Rouse}
U_p = \frac{1}{N} \sum_{n=1}^N R_n \cos\left[  \frac{(2n-1)p\pi}{2N}\right]\,,
\end{equation}
where $p=0,1\cdots,N-1$ is the mode number and $U_p$ is the mode amplitude.
(The choice of cosines automatically takes care of the
modified equations for $R_1$ and $R_N$.)
The Rouse coordinates are decoupled and evolve according to~\cite{doi}
\begin{equation}\label{eq:Rouse_mode}
\zeta_p \frac{d U_p}{dt} = -\kappa_{p} U_p + W_p\,,
\end{equation}
where $\kappa_{p}=8N\kappa\sin^2\left(\frac{p \pi}{2N}\right)$,
and $\zeta_p=2N\zeta$ for $p\ge 1$, and $\zeta_0=N\zeta$. The
noise $W_p$ has again zero mean and correlations of
\begin{equation}\label{eq:FDcondition}
\langle W_{p} \left(t\right) W_{p'}\left(t'\right) \rangle =
2k_{B}T \zeta_p\delta_{p,p'}\delta(t-t')\,.
\end{equation}
Note that the $N-1$ {\em internal modes} for $p\neq 0$ behave as particles
tethered by a harmonic spring, while the center of mass (CM), corresponding
to $p=0$, freely diffuses.

The linear Eqs.~\eqref{eq:Rouse_mode} are readily solved starting from
any initial condition, and the  probabilities for $\{U_p\}$ are Gaussians,
with a time-dependent mean set by initial conditions, and a variance
\begin{equation}\label{eq:sigma_p}
\begin{aligned}
\sigma_p^2 &=\frac{k_BT}{\kappa_p}(1-{\rm e}^{-2t/\tau_p}),\ {\rm for\ }p\ge 1\,, \\
\sigma_0^2 &= 2D_{\textsc{cm}}t\,.
\end{aligned}
\end{equation}
The equilibration (relaxation) times of the internal modes are
$\tau_p=\zeta_p/\kappa_p$,
while the diffusion constant  is $D_{\textsc{cm}}=k_BT/N\zeta$.
There is clearly a hierarchy of relaxation times: the shortest timescale
$\tau_{N-1}\approx \zeta/(4\kappa)$ is half of the time
$\tau_{\rm s}=\zeta/(2\kappa)$ during which a {\em free} monomer diffuses
a mean squared distance between the adjacent monomers,
$b^2=k_BT/\kappa$.  For the CM, we can define a characteristic time
$\tau_0\equiv b^2N/D_{\textsc{cm}}=N^2\zeta/\kappa$ associated with
diffusing over the size of the polymer.
This is of the same order as the longest internal relaxation time $\tau_1$;
for long polymers $\tau_0/\tau_1\approx\pi^2$.

By inverting Eq.~\eqref{eq:def_Rouse} one can also follow the position of
a specific (``tagged") monomer, as
\begin{equation}\label{eq:def_inv}
R_n =U_0+2\sum_{p=1}^{N-1} U_p  \cos\left[  \frac{(2n-1)p\pi}{2N}\right]\,.
\end{equation}
This equation becomes particularly simple for the central monomer
$c=(N+1)/2$ (assuming odd $N$), as
\begin{equation}\label{eq:R_c}
R_c =U_0+2\sum_{k=1}^{(N-1)/2} (-1)^kU_{2k}\,.
\end{equation}
Since each term in the above sum is (independently) Gaussian distributed,
so is $R_c$, with variance
\begin{equation}\label{eq:VarR_c}
{\rm var}(R_c)=\sigma^2_0+4\sum_{k=1}^{(N-1)/2}\sigma^2_{2k}\,.
\end{equation}
Utilizing Eq.~\eqref{eq:sigma_p}, we can now distinguish between three regimes:
\begin{enumerate}
\item For short times $t\ll \tau_{\rm s}$, we can expand the exponential in
Eq.~\eqref{eq:sigma_p} and obtain $\sigma_p^2\approx k_BT t/(N\zeta)$
independent of $p$. The sum in Eq.~\eqref{eq:VarR_c} then leads to
${\rm var}(R_c)\approx 2D_{\rm s}t$, with a monomer diffusion constant of
$D_{\rm s}=k_BT/\zeta$.
\item For very long times $t\gg \tau_1$ all the internal modes saturate to a variance
that is independent of $t$. In this regime the additional time dependence comes from
the first term in Eq.~\eqref{eq:VarR_c} corresponding to the slow diffusion of the center of mass.
\item The most interesting regime is for intermediate times, with
$\tau_{\rm s}\ll t\ll \tau_0$, where only the terms for which $2t/\tau_{2k}>1$
will contribute significantly to Eq.~\eqref{eq:VarR_c}. Focusing on the
corresponding modes, we obtain
\begin{equation}\label{eq:approxVar_R_c}
{\rm var}(R_c)\approx 4\int_{k=k_{\rm min}}^{N/2}\frac{k_BT}{\kappa_{2k}}dk\,,
\end{equation}
where $k_{\rm min}$ is determined from the relation $\tau_{2k_{min}}=2t$.
Using the usual expression for $\tau_p$ of the Gaussian polymer
one can immediately see that ${\rm var}(R_c)\sim t^{1/2}$.
This interval clearly exhibits anomalous diffusion due to the collective behavior
of the modes; its size can be made arbitrarily long by letting $N\to\infty$.

\end{enumerate}

\subsection{Generalized modes with variable exponent}\label{sec:mod}

In Ref.~\cite{kkgauss} the tagged monomer was used as a prototype of subdiffusion.
Of course, as described so far the anomalous dynamics of the tagged monomer is
characterized by the exponent $\alpha=1/2$.
It is possible to modify Eq.~\eqref{eq:Rouse_mode} in different ways so as to
produce dynamics for any value of $\alpha$.
We shall do so by considering power-law dependences of the friction coefficients
$\zeta_p$ on the mode-index $p$, as motivated by the following two physical models:

\begin{enumerate}
\item Zimm analyzed the motion of a polymer in the presence of hydrodynamic flows,
which result in interactions that decay (in three dimensions) as an inverse
distance between monomers. These interactions do not change the probability
distribution in configuration space (which is still governed by the equilibrium
Boltzmann weight), but do modify the relaxation times.
Zimm showed~\cite{zimm} that the resulting dynamics can be {\em approximately}
described by Eq.~\eqref{eq:Rouse_mode}, but with $\zeta_p\propto p^{1/2}$.
We can now work through the same steps as in the previous section, and
find anomalous diffusion for the tagged monomer, but with exponent $\alpha=2/3$.

\item  Equation~(\ref{eq:Hamiltonian}) can also be regarded as the energy of
a fluctuating stretched line, with $\{R_n\}$ indicating the heights above a
baseline \cite{krug,kardar_book}.
If the line separates two phases of fixed volume, the sum $\sum_n R_n$ must remain
unchanged under the dynamics.
Such {\em conserved dynamics} are mimicked by Eq.~\eqref{eq:Rouse_mode},
but with $\zeta_p\propto p^{-2}$ \cite{kardar_book}.
As long as the noise correlations satisfy Eq.~\eqref{eq:FDcondition} the equilibrium
distribution remains unchanged, but the dynamics is slowed down, such that the
fluctuations of a given coordinate now evolve with exponent $\alpha=1/4$.
\end{enumerate}

Motivated by the above examples, we consider dynamics according to Eqs.~(\ref{eq:Rouse_mode})
and (\ref{eq:FDcondition}) with generalized $\zeta_p$ given by
\begin{equation}\label{model_friction_coefficient}
\begin{aligned}
\zeta_p &= 2C N \zeta \left( \frac{p}{N} \right)^{2-z},\ {\rm for\ }p\ge 1\,, \\
\zeta_0 &=  \zeta N^{z-1}\,.
\end{aligned}
\end{equation}
The choice of exponent $z$ leads to time-scales
\begin{equation}\label{eq:tau_p}
\tau_p=\frac{C\zeta(p/N)^{2-z}}{4\kappa\sin^2(p\pi/2N)}
\approx C\frac{\zeta}{\kappa\pi^2}\left(\frac{N}{p}\right)^{z}\,,
\end{equation}
where the last approximation is valid for $p\ll N$.
The longest internal mode scales with the number of degrees of freedom as $N^z$.
This is the conventional notation for the dynamic exponent of a fluctuating line,
or a free-field theory, but somewhat different from that used to denote the
relaxation of polymers.
The Rouse, Zimm, and conserved models correspond to $z=2,~3/2$ and 4, respectively.
The dimensionless constant $C=2\zeta_1/\zeta_0$ is somewhat arbitrary. It
defines the ratio between the time characterizing the diffusion of
the CM and the relaxation time of the slowest internal mode.
We chose it in a way that for very short times the motion of $R_c$  has a
diffusion constant of $D_{\rm s}=k_BT/\zeta$ irrespective of $z$. For large
$N$ this leads to $C=1/(z-1)$.
We would like to stress that Zimm dynamics, as well as other physical systems producing anomalous diffusion, are only {\it approximately} described by Eq.~\eqref{eq:Rouse_mode} with length-scale dependent
friction constants. However, we employ Eqs.~\eqref{eq:Rouse_mode}
and \eqref{model_friction_coefficient} as the (exact) definition of
our mathematical model for anomalous diffusion with tunable exponents.

Focusing on the coordinate $R_c$ (the tagged monomer), we observe that it executes
normal diffusion with diffusion coefficient $D_{\rm s}$ for times $t\ll \tau_{\rm s}=\zeta/2\kappa$.
For very long times $t\gg \tau_0=(\zeta/\kappa)N^{z}$  it again performs regular diffusion
but with a much smaller center of mass diffusion coefficient  of $D_{\rm s}/N^{z-1}$.
At intermediate times $\tau_{\rm s}<t<\tau_0$, fluctuations of $R_c$ are influenced by
the   $p$-dependent relaxation times in Eq.~\eqref{eq:tau_p}, and evolve anomalously
as ${\rm var}(R_c)\propto t^\alpha$. The exponent $\alpha$ can be obtained simply by
noting \cite{KBG} that at times of order $\tau_0$, ${\rm var}(R_c)$ should be
similar to typical equilibrium (squared) size of the polymer, which grows as
$Nk_BT/\kappa$. Equating these two quantities immediately yields
\begin{equation}\label{eq:anom_dif}
{\rm var}(R_c)= K b^{2}\left(\frac{D_{\rm s}t}{b^2}\right)^\alpha\,,
 \end{equation}
where $K$ is a dimensionless prefactor, and
\begin{equation}\label{anomalous_exponent}
\alpha=\frac{1}{z}\,.
\end{equation}
This result can also be directly obtained from Eqs.~\eqref{eq:approxVar_R_c}
and \eqref{eq:tau_p}.

In Ref.~\cite{krug} an alternative strategy is employed for obtaining a
tunable exponent, namely by scaling the ``spring constant" in
Eq.~\ref{eq:Rouse_mode} as $\kappa_p^{2/z}$, while leaving the friction
coefficients $\zeta_p$ unchanged. Without a corresponding scaling of the
noise amplitudes $W_p$, the steady state probability is now also modified.
The simulations of Ref.~\cite{krug} are actually obtained by evolving the
Langevin equations in real space (as opposed to Fourier mode evolutions
performed in our current work). This necessitates generalizing the
interactions in Eq.~\ref{eq:Rouse_Langevin} to further neighbors, and/or
generating correlated noise in real space. Nevertheless, for $z=2$ the two
approaches should be identical -- corresponding to Gaussian polymers --
and direct comparison should be possible.

\subsection{Numerical implementation}\label{subsec:numerics}

\begin{figure}
\includegraphics[height=6cm]{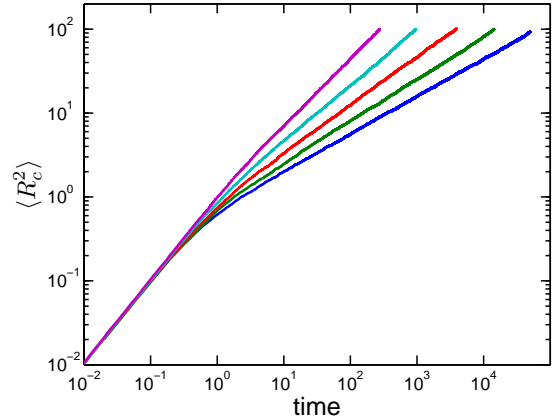}
\caption{(Color online) Mean squared position of the central monomer
(in units of $b^2$) as a function of time (in units of $\tau_{\rm s}$)
for a polymer of length $N=257$.  The curves correspond (left to right)
to  $z=1.25,~1.5,~1.75,~2$, and $2.25$. Each curve is an
average over 10,000 realizations.}
\label{fig:var_vs_t}
\end{figure}

To verify the dependence of the anomalous exponent on $z$, we simulated the
dynamics of a chain of $N=257$ monomers by numerically solving the Langevin
equation for each of the modes.  At the beginning of each simulation we
equilibrated the polymer by randomizing the initial mode amplitudes, and
positioned the central monomer ($c=129$) at the origin. The position $R_c$
of the central monomer was evolved by numerically integrating the Langevin
Eqs.~\eqref{eq:Rouse_mode} with the  Smart Monte Carlo method~\cite{rdf},
followed by transforming the mode amplitudes $U_p$ to the monomer position
space, at each time step. Figure~\ref{fig:var_vs_t} illustrates the results
for several values of
$z$. The averages of $R_c^2$ were calculated over 10,000 independent
simulations. For each value of $z$ the times are at least an order of
magnitude shorter than the slowest relaxation mode of the polymer for
that $z$. For $t<\tau_{\rm s}$ the curves coincide: in that region the
particles perform normal diffusion with diffusion constant $D_{\rm s}$
which is independent of $z$. For $t>\tau_{\rm s}$ we can clearly observe
a pure power law growth with exponent $\alpha=1/z$ confirming
Eq.~\eqref{anomalous_exponent}. Using our model it is possible to get very
slow dynamics $\alpha \rightarrow 0$ by taking $z \gg 1$, or to reproduce
$\alpha=1$ by setting $z=1$.

We also verified that the probability density of the distribution of $R_c$
is a Gaussian. The Langevin equation describing each $U_p$ can be solved
analytically, and consequently the parameters of the Gaussian distribution
for $R_c$ (its mean and variance) are known. Thus, the numerical results
presented in Fig.~\ref{fig:var_vs_t} were anticipated, and primarily served
to evaluate the accuracy of the numerical procedure. In the following
sections we will use the same procedure for results that cannot be found
analytically.

\section{Results}

The behavior of a particle near absorbing boundaries may reveal aspects
of the dynamics not apparent in the scaling of the mean square displacement.
Indeed, consideration of absorption of a monomer in a simple Gaussian
polymer~\cite{kkgauss} have provided novel insights into the differences
between different forms of anomalous dynamics with $\alpha=1/2$.
In our numerical studies, we implemented anomalous diffusion as described
in Sec.~\ref{subsec:numerics}, but with {\it only the tagged monomer
interacting with the absorbing boundaries}. For the case of a single
absorbing boundary, for each $z$, the starting position $x(0)$
of the tagged monomer was at a distance $8b$ away from the absorbing
boundary, while for the case of two absorbing boundaries the monomer was
placed between them at a distance $8b$ from either one. The numerical
procedure imposes both lower and upper limits on the relevant times:
\begin{enumerate}
\item If the tagged monomer is initially located at a distance $x(0)$ from an
absorbing boundary, a sufficiently long time is required for the
probability density to be influenced by absorption. Since the typical
squared distance travelled by anomalous diffusers is given by
Eq.~\eqref{eq:anom_dif}, the absorption probability becomes significant
after a time (disregarding the dimensionless prefactor $K$)
\begin{equation}\label{eq:T}
T=(b^2/D_{\rm s})[x(0)/b]^{2/\alpha}=2\tau_{\rm s}[x(0)/b]^{2/\alpha}\,.
\end{equation}
In our simulations with $x(0)=8b$, and for $z=1.25$ ($2.25$) this
leads to $T\approx 360\tau_{\rm s}$  ($2.3\times 10^4\tau_{\rm s}$).
Alternatively, one can find directly from Eq.~\eqref{eq:VarR_c}, that
for $z=1.25$ ($2.25)$ the quantity $\langle R_c^2\rangle$ reaches $64b^2$
at time $T'\approx 160\tau_{\rm s}$ ($2.3\times 10^4\tau_{\rm s}$).
(Graphically, $T'$ can be found simply from Fig.~\ref{fig:var_vs_t} as
the time corresponding to $64b^2$.)
The fact that $T'$ is not strictly proportional to $T$, is
related to the $z$-dependence of $K$, since $z$ enters
Eqs.~\eqref{model_friction_coefficient} through the constant $C$.
\item Anomalous diffusion is expected for times significantly shorter than
the longest relaxation time  $\tau_0\approx\tau_{\rm s}N^{z}$; for
$z=1.25$ ($2.25$)  and  $N=257$, we estimate
$\tau_0\approx 10^3\tau_{\rm s}$ ($2.6\times 10^5\tau_{\rm s}$).
[Since $\tau_1\approx\tau_0/[(z-1)\pi^2]$ (see Eq.~\eqref{eq:tau_p}),
the corresponding values of the slowest internal modes are
$400\tau_{\rm s}$ ($2.1\times 10^4\tau_{\rm s}$.)]
In the presence of a single absorbing boundary our simulation
times for $z=1.25$ ($2.25$) were shorter than
$1.6\times 10^3\tau_{\rm s}$ ($1.6\times 10^4\tau_{\rm s}$),
while for the case of two absorbing boundaries they were
shorter than $600\tau_{\rm s}$ ($6.5\times 10^3\tau_{\rm s}$).
These numbers indicate that most of our simulations stayed within the
anomalous diffusion regime. To verify this point
we performed simulations for $N=65,~129$, and $257$ for $z=2$
and observed the convergence of their absorption time
distributions, which indicates that we are in the $N$-independent
regime. In the following we report only the results for the central
monomer $c=129$ of the chains with $N=257$.
\end{enumerate}

\subsection{Absorption time distribution}\label{sec:abs}

\subsubsection{Single absorbing boundary}

The problem of a particle preforming normal diffusion in the presence of
absorbing  boundaries has been described in detail by
Chandrasekhar~\cite{chandra}. It can be cast as a simple (linear) diffusion
equation for the evolving probability density, with vanishing boundary
conditions at the absorbing points. In the presence of one absorbing
boundary in 1D, an elegant solution is found by the method of
images~\cite{chandra,redner}, i.e.\ by subtracting from the Gaussian
solution describing the probability density in the absence of absorption,
a similar Gaussian centered at the ``mirror image position" with respect
to the absorbing boundary. At times shorter than $T=x^2(0)/D$,  where $D$
is the diffusion constant of the particle and $x(0)$ is its initial distance
from the boundary, the particle does not feel the absorbing boundary.
For $t\gg T$ the survival probability (obtained by integrating the
solution over the allowed interval) scales as $S(t)\sim t^{-1/2}$, while
the absorbtion PDF behaves as $Q(t)=-dS/dt\sim t^{-3/2}$.  We note that the
mean absorbtion time is infinite, since the particle can survive indefinitely
by diffusing away from the absorbing  boundary.

%%%%%%%%%%%%%%%%%%%%%%%%
\begin{figure}
\includegraphics[height=6cm,clip]{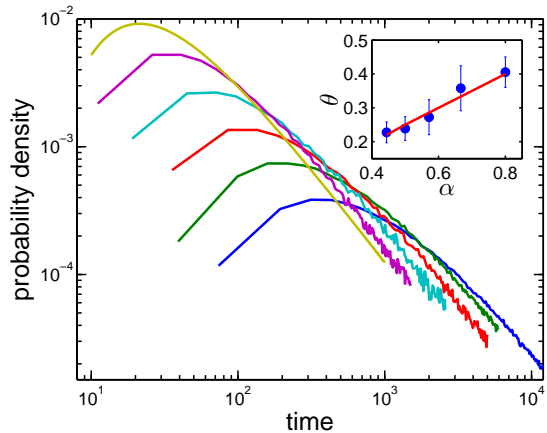}
\caption{(Color online) Logarithmic plot of absorption probability distribution
as a function of time (in units  of $\tau_{\rm s}$) of the central monomer
of a $N=257$ polymer, in the presence of an absorbing boundary at a distance $8b$
from the initial position of the monomer. The leftmost curved
depicts the result for normal diffusion of a single particle. The rest
of the curves correspond (left-to-right) to $z=1.25,~1.5,~1.75,~2,$ and $2.25$.
For each $z$ value, 100,000 independent runs were performed.
The inset depicts the value of $\theta$ obtained from the
slopes of these graphs as a function of $\alpha$. The continuous
line depicts the relation between these exponents proposed in
Ref.~\cite{brd} (see text).}
\label{fig:Q1wall}
\end{figure}
%%%%%%%%%%%%%%%%%%%%%%%%%

Figure~\ref{fig:Q1wall} depicts on a logarithmic
scale the PDFs of the absorption $Q(t)$ of the tagged monomer
initially located at distance $x(0)=8b$ from a single absorbing
boundary for  different values of $z$. (For comparison,
results for normal diffusion of a single particle are also shown.)
Absorption is negligible at short times, but gradually increases to a
maximum at a time significantly smaller than $T$ defined by Eq.~\eqref{eq:T}.
As in the case of normal diffusion, it is generally accepted that the
absorption PDF decays as a power law $Q(t)\sim t^{-1-\theta}$ at long
times; $\theta=1/2$ for regular diffusers while any $\theta\le 1$ leads
to a diverging mean absorption time. We attempted to extract the exponent
$\theta$ from the slopes of the curves in the logarithmic plot in
Fig.~\ref{fig:Q1wall} at large values of $t$. While $10^4$ independent runs
were performed to  obtain each of the curves, only a small
fraction of diffusers survived to times significantly
longer than the position of the maximum. Nevertheless, for each
$z$ we have reasonably accurate results for 1.5 decades
beyond the time of maximal $Q(t)$. Only the second
half of this interval (on a logarithmic scale) is a
straight line, and it was used to evaluate $\theta$.
There are thus significant statistical errors in the estimates for $\theta$
as depicted by the error bars in the inset of Fig.~\ref{fig:Q1wall}.

Studies of continuous time random walks
using the fractional Fokker-Planck equation~\cite{brd} obtain
a simple relation $\theta=\alpha/2$ for $0<\alpha<2$.
This relation is  depicted by the continuous line in the inset.
Although there is no sound theoretical foundation for applying this
relation to the collective anomalous diffusion of our model, we note
that there is some correspondence between the line and the
measured exponents. In Ref.~\cite{krug} an alternative relation
$\theta=1-\alpha/2$ is proposed, based on considerations of fractional
Brownian motion. Neither the values, nor the trend in this relation
are consistent with the results in Fig.~\ref{fig:Q1wall}. We are puzzled
by this discrepancy as Ref.~\cite{krug} also provides numerical support
for this relation, and the method used (although somewhat distinct in
general) should essentially coincide with ours for the case of $z=2$
($\alpha=1/2$). We are reluctant to make a definite statement regarding
this exponent, as in addition to the statistical errors (error bars in
the inset in Fig.~\ref{fig:Q1wall}) there are uncertainties due to
possible systematic errors: The measurements of $\theta$ are performed
for times one order of magnitude larger that of the maximum of $Q(t)$.
These times  do not significantly exceed $T$ or $T'$, and are possibly
even shorter than the latter, especially for larger values of $z$.
Getting rid of such possible  systematic errors requires significantly
larger $N$ and more statistical samples, but is clearly needed to
resolve this discrepancy.

\subsubsection{Two absorbing boundaries}\label{subsec:abs2}

\begin{figure}
\includegraphics[height=6cm]{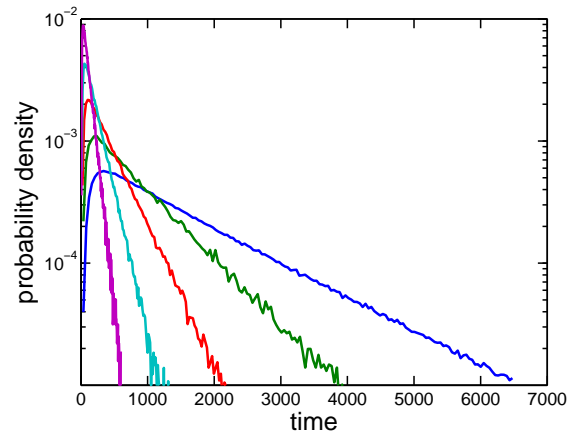}
\caption{(Color online) Semi-logarithmic plot of the absorption probability density
of a tagged monomer as a function of time (in units  of $\tau_{\rm s}$)
for the central monomer in a $N=257$ polymer, in the interval between
two absorbing boundaries at a distance $16b$. The monomer starts at
the midpoint between the boundaries. Each curve is the result of 100,000
independent simulations. The different curves correspond (left to right) to $z=1.25,~1.5,~1.75,~2$, and $2.25$.}
\label{fig:Q2wall}
\end{figure}

We also considered a tagged monomer confined in the interval between
two absorbing boundaries separated by $16b$. Initially the particle is placed
half-way between the two boundaries, i.e.\ at a distance $8b$ from each.
Figure~\ref{fig:Q2wall} illustrates on a semi-logarithmic plot, the
absorption probability $Q(t)$ for different values of $z$. The distributions
have the same general shape as before, with absorption probability rising
with time to a maximum. However, the fall off at long times appears to be
exponential. The straight lines in the semi-logarithmic plots clearly rule
out the stretched  exponential decay characterizing other forms of anomalous
diffusion~\cite{nechaev}. They also bear no resemblance to the power law
decay (with diverging mean absorption time) expected for subdiffusion
described by the fractional Fokker-Plank equation~\cite{fpinf}.
The mean absorption time and the typical decay time (as determined
from exponential decay) are practically indistinguishable for
each $z$. They are, however, by more than an order of magnitude
shorter than the typical time $T$ defined by Eq.~\eqref{eq:T} or $T'$.

%%%%%%%%%%%%%%%%%%%%%%%%%
%%%%%%%%%%%%%%%%%%%%%%%%%

%%%%%%%%%%%%%%%%%%%%%%%%%%%%%%%%%%%%%%%%%%%%%%%%%%%%%%%%%%%%%%%%%%%%%%%%%%%%%%
\subsection{Long-time distribution of particle position}\label{sec:pdfs}
\subsubsection{Single absorbing boundary}\label{subsec1:pdfs}

Let us now consider the dependence on the coordinate $x$ of the PDF $p(x,t|x(0))$,
starting at an initial position $x(0)$ from a single absorbing boundary at $x=0$.
As noted before, for a regularly diffusing particle this PDF is obtained by
the method of images as the difference between Gaussians centered at $\pm x(0)$,
whose width grows as $\ell(t)=\sqrt{D_{\rm s}t}$. Expanding this solution close
to the boundary, we find $p(x,t|x(0))\approx xx(0)/\ell(t)^3$, i.e. the PDF vanishes
linearly with $x$ in the vicinity of the boundary. It is tempting to generalize
this result to anomalous diffusion, by simply replacing the scaling form of the
width by $\ell(t)=b^{1-\alpha}(D_{\rm s}t)^{\alpha/2}$, and again concluding a
linear behavior with $x$ albeit with a different dependence on $t$. However, the
method of images relies on the absence of memory in the motion~\cite{chandra},
which is not correct for our non-Markovian processes. Indeed in a previous study
we observed that for the case of $\alpha=1/2$ the vanishing of the PDF near an
absorbing boundary is faster than linear. (A similar problem occurs when a
diffuser performs L\'{e}vy flights, since the absorbing boundary is no longer a
turning point of the trajectory \cite{zoia,chechkin}.) We shall assume that for
$t\gg T$ (Eq.~\eqref{eq:T}) the behavior near the boundary can be
described by  $p(x,t|x(0))\sim x^\phi$.

\begin{figure}
\includegraphics[height=6cm]{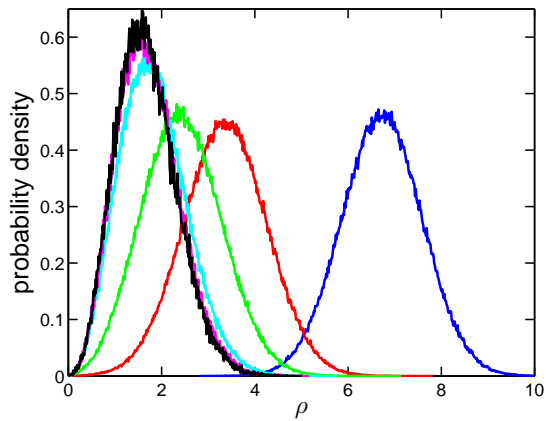}
\caption{(Color online) Probability density function of the central monomer $c=129$ in
a polymer with $N=257$ monomers in the presence of one absorbing boundary.
The monomer is initially located at a distance $8b$ from the absorbing
boundary. The horizontal axis is in the scaled variable $\rho=x/\ell(t)$.
The graphs correspond to different times (right to left)
$t/\tau_{\rm s}=2$, 32, 128, 1024, 3500, 6000, and obtained
from 100,000 independent runs.}
\label{fig:collapse}
\end{figure}

In the long time limit we expect $\ell(t)$ to be the only relevant length
in the problem. However, the initial separation, $x(0)$, from the
absorbing boundary is another length scale, which may become irrelevant
only for $\ell(t)\gg x(0)$. To check this assumption we plotted the PDF of
unabsorbed particles in terms of the scaled variable $\rho\equiv x/\ell(t)$
for $N=257$ polymers. Figure \ref{fig:collapse} depicts a sequence of PDFs
for $z=2$ ($\alpha=1/2$) at several times. For short times, i.e. when
$x(0)\gg\ell(t)$ the maximum of the PDF remains centered close to $x(0)$,
and therefore, its center appears near
$\rho=x(0)/\ell(t)$, and moves to smaller values of $\rho$
with increasing $\ell(t)$. Indeed this process is clearly seen in
Fig.~\ref{fig:collapse} for the three graphs representing the short times,
with their maxima moving to the left as $t^{-1/4}$. Graphs
corresponding to the two largest times almost coincide representing the
final stable shape. Note that this behavior appears in the same range
as the apparent power law behavior of $Q(t)$ in Fig.~\ref{fig:Q1wall},
and above $T$ defined in Eq.~\eqref{eq:T}. To verify the stability of
the scaled PDFs it is desirable to study even larger times. Unfortunately,
the quality of the graphs deteriorates since the number of surviving
diffusers becomes very small.

%%%%%%%%%%%%%%%%%%%%%%%
%%%%%%%%%%%%%%%%%%%%%%%%

%%%%%%%%%%%%%%%%%%%%%%%
\begin{figure}
\includegraphics[height=6cm]{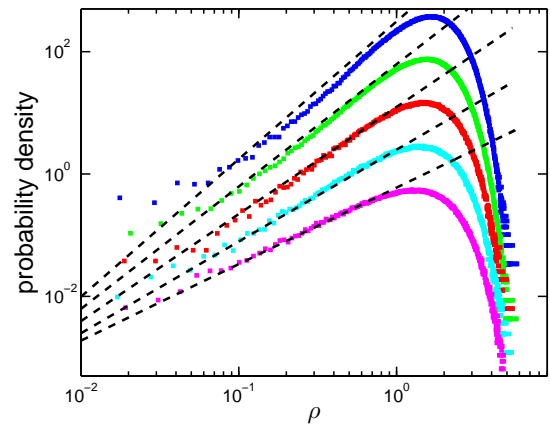}
\caption{(Color online) Probability density function of the central particle ($c=129$)
in chains of $N=257$ monomers in the presence of one absorbing boundary.
The horizontal axis is in the scaled variable $\rho=x/\ell(t)$ (see text).
The different curves correspond to $z=1.25,~1.5,~1.75,~2,$ and $2.25$ (bottom
to top). The graphs are shifted vertically for clarity with increasing $z$
values by a factor of 5. The dashed lines have slopes $\phi=1/\alpha$.
Each curve was obtained from 100,000 independent runs.}
\label{fig:Px1wall}
\end{figure}
%%%%%%%%%%%%%%%%%%%%%%%%

Figure~\ref{fig:Px1wall} depicts on a logarithmic scale the PDF of the
particle position in terms of the scaled variable $\rho=x/\ell(t)$
for several values of $z$. Since the evaluation of the probabilities
is performed at large times, when only a small fraction of the initial
100,000 samples survives, the statistical fluctuations are significant.
The figure was obtained by using rather large bins, which poses a problem
for a function fast approaching zero. In Fig.~\ref{fig:Px1wall} about
five leftmost points of each graph containing small numbers of events
should be disregarded due to statistical uncertainty and more importantly
due to the distortion caused by the bin sizes. These effects severely
limit the accuracy with which we can determine the exponent $\phi$. As a
guide to the eye we have added straight lines with slopes $\phi=1/\alpha$,
which seem to provide a fair approximation of the slopes  in the range
$0.1<\rho<1$. We note that such a form of $\phi$ describes the  behavior
near an absorbing wall for L\'{e}vy flights~\cite{zk}, as well as for
diffusion described by a fractional Laplacian~\cite{zoia} between two
absorbing boundaries. Again, as emphasized before, one should beware of
possible systematic errors, as the times for which the PDF is measured
are not particularly long.

Recently, Zoia {\em et al.} argued~\cite{zrm} that under rather general
assumptions there is a relation between the anomalous diffusion exponent
$\alpha$, the exponent $\theta$ governing the tail of the absorption PDF,
and the boundary exponent $\phi$, given by  $\phi=2\theta/\alpha$. Relying
on $\theta=1-\alpha/2$ (Ref.~\cite{krug}), they thus obtain
$\phi=2/\alpha -1$ which is larger than the estimates from
Fig.~\ref{fig:Px1wall}. On the other hand, using our fits with
$\theta\approx\alpha/2$ would yield $\phi\approx 1$, which is smaller
than our data indicates.

%%%%%%%%%%%%%%%%%%%%%%%%%%%%%%%%%%%%%%%%%%%%%%%%%%%%%%%%%%%%%%%%%%%%%%%%%%%%%%%%%%%%%%%
\subsubsection{Two absorbing boundaries}\label{subsec2:pdfs}

In Sec.~\ref{subsec:abs2} we noted that with two absorbing boundaries the
absorption probability eventually decays exponentially.
Indeed, the time dependent PDF for a normal diffuser between two absorbing
boundaries is represented by a sum of spatial sinusoidal modes
(eigenfunctions of the Laplacian operator) multiplied by functions of time
which are pure exponentials. At large times only the lowest harmonic
corresponding to the slowest decay, survives. Thus for  normal diffusion
in the interval between two absorbing boundaries at
$X_{b1,2}=\pm 8b$, the spatial probability at long times behaves as
$\sim \cos(\pi x/16b)$, again vanishing linearly at the endpoints.
In Ref.~\cite{kkgauss} it was demonstrated that for $\alpha=1/2$,
at times significantly larger than the mean absorbtion time, the
normalized PDF of positions of the surviving tagged monomer has a stable
shape different from a cosine.

\begin{figure}
\includegraphics[height=6cm]{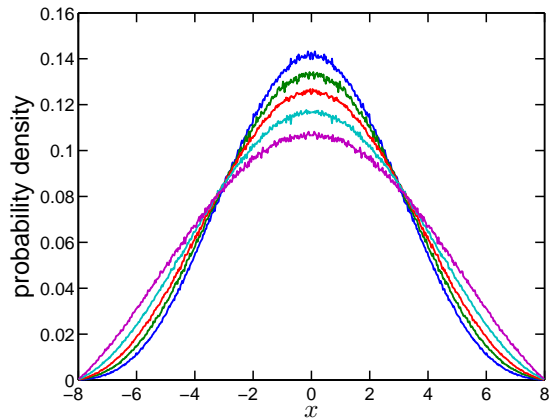}
\caption{(Color online) Probability density function of the central particle in a
chain with $N=257$ monomers with absorbing boundaries at
$X_{b1,2}=\pm 8b$ as a function of position measured from the center
of the interval (in units  of $b$)
for $z=1.25,~1.5,~1.75,~2$, and $2.25$ (broad to narrow). Each graph is the result
of 100,000 independent runs.}
\label{fig:Px2wall}
\end{figure}

We performed a detailed study of spatial dependence of the PDF of the
surviving anomalous walker between two absorbing boundaries for several
values  of $z$. The properly normalized PDF of surviving monomers
progresses from a Gaussian with variance growing linearly in time (for
$t\ll\tau_{\rm s}$), to a Gaussian with variance increasing as
$t^\alpha$ at intermediate times, before settling down to a stable shape
beyond the point  of maximum of $Q(t)$ in Fig.~\ref{fig:Q2wall}.
Figure~\ref{fig:Px2wall} depicts these stable shapes for several values
of $z$. Note the non-linear behavior of the curves near the boundaries.
Interestingly the boundary exponent appears to approach $\phi=1$ as
$z\to 1$. This is indeed the expected behavior for a normal diffuser,
although we note that our diffusers in the limit $z=1$ still reflect
the collective behavior of many modes.

%%%%%%%%%%%%%%%%%%%%%%
%%%%%%%%%%%%%%%%%%%%%%%

\begin{figure}
\includegraphics[height=6cm]{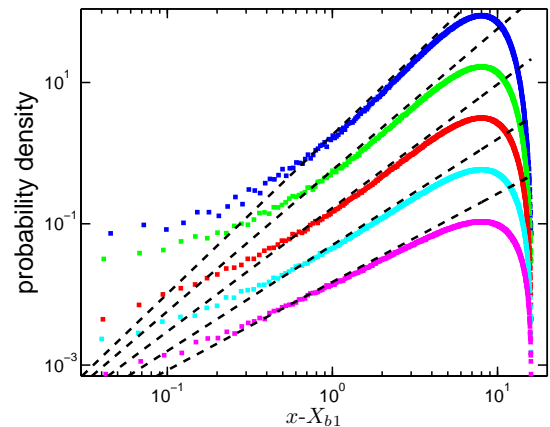}
\caption{(Color online) Probability density function of the central monomer in a
chain with $N=257$ particles in the presence of absorbing boundaries at
$X_{b1} = -8b$ and $X_{b2} = 8b$, for $z=1.25,~1.5,~1.75,~2$, and $2.25$
(bottom to top). Each curve is the result of 100,000 independent
runs. The curves are shifted vertically for clarity, by a factor of
5 at each increasing $z$. The dashed lines have slopes
$\phi=1/\alpha$.}
\label{fig:Px2wall_log}
\end{figure}

To better display the behavior of these stable functions near the
boundaries, in Fig.~\ref{fig:Px2wall_log} we plot them on a logarithmic
scale as a function of a distance from one edge. The results are distorted
not only for the reasons mentioned in Sec.~\ref{subsec1:pdfs}, but also
because of the smearing caused by the finite time step of the algorithm
(causing a typical step size of each monomer). As in the case of a single
absorbing boundary, the dashed lines with slopes $\phi=1/\alpha$ are
drawn to guide the eye. Since the curves
have similar shapes and approximately follow a power law near the
boundary, we attempted to collapse various curves by raising them
to power $2\alpha$ and normalizing them, but this procedure
{\em did not} result in a good data collapse.

%%%%%%%%%%%%%%%%%%%%%%%%
%%%%%%%%%%%%%%%%%%%%%%%%%

%%%%%%%%%%%%%%%%%%%%%%%%%%%%%%%%%%%%%%%%%%%%%%%%%%%%%%%%%%%%%%%%%%%%%%
\section{Discussion}
\subsection{Comparison with translocation}\label{sec:transloc}

As we were initially lead to this subject in connection with polymer
translocation, it is fitting to conclude by returning to this issue.
Translocation, the passage of a polymer through a pore in a membrane,
is an important process that has been studied extensively in
the last decade~\cite{park,muthu,Lubensky,chern,ckk,kasian,akeson,mel}.
Phages, for example, invade bacteria
by taking advantage of existing channels in bacterial membranes
to translocate their DNA/RNA inside~\cite{dreis}.
In theoretical models, it is convenient to study a translocation
coordinate $s$ which denotes the number of monomers $s$ on one side of the pore.
The dynamics of this coordinate is anomalous:
If we assume that the translocation time is
of the order of polymer relaxation time $\tau_0$~\cite{ckk},
then the variance of $s$ will increase with time as $t^{\alpha'}$
with $\alpha'=2/(2\nu+1)$ (Rouse dynamics). (For a self-avoiding polymer
diffusing in 2D, $\nu=3/4$ and $\alpha'=0.8$.)
We note that the actual value of the exponent $\alpha'$ also involves
factors not explicitly related to the polymer dynamics: (i) that
self-avoiding effects expand the equilibrium size of the polymer in the
physical space; and (ii) that the relevant variable represents a 1D
coordinate in the {\it internal} space of monomer numbers.
While the expression for the actual exponent has been
supported numerically in some
studies~\cite{ckk,luo2d,huopa2d,chat,luo2dand3d}, and disputed in
others~\cite{panja,dubb,milchev}, the anomalous nature
of dynamics is not in question.

To simplify the process, numerical implementations frequently begin by
inserting half of the polymer into the pore (i.e. $s(t=0)\approx N/2$)
and allowing the polymer to diffuse until either of its ends ($s=1$ or $s=N$)
leaves the pore.  This closely resembles the motion of anomalous diffuser in
Sec.~\ref{subsec2:pdfs} between two absorbing walls, and in a previous
work~\cite{kkgauss} for $\alpha=1/2$. We are now in position to make a more
meaningful comparison by choosing a value of $z$ that reproduces the observed
exponent for anomalous dynamics of $s(t)$. Specifically, recently Chatelain
{\em et al.} \cite{chat} preformed high accuracy simulations of two-dimensional
translocation of a self-avoiding polymer. Relevant conclusions from this work
are: (a) At short times the distribution of $s$ is almost an exact Gaussian
whose variance increases in time with exponent $\alpha'\approx0.8$.
(b) For times larger than the typical translocation time  the distribution
of translocation time decays exponentially, as in $Q(t)$ in the current work.
(c) For times significantly larger than the mean translocation time the PDF
of the surviving translocation coordinate qualitatively resembles those in
Fig.~\ref{fig:Px2wall} of Sec.~\ref{subsec2:pdfs}.

%%%%%%%%%%%%%%%%%%%%%%%
\begin{figure}
\includegraphics[height=6cm]{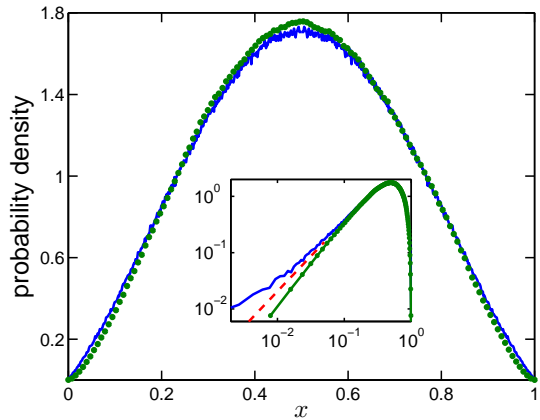}
\caption{(Color online) The dots connected by a line depict the normalized PDF of the
translocation coordinate, at times significantly exceeding the mean translocation
time, for a two-dimensional self-avoiding polymer of length $N=128$ that did
not yet translocate  (from Ref.~\protect{\cite{chat}}). This is compared to
the normalized PDF of a central particle (in a chain of $N=257$)
performing anomalous  diffusion controlled by exponent $z=1.25$
(i.e. $\alpha=0.8$), moving between two absorbing boundaries.
For the purpose of comparison the range of the translocation
coordinate and the range of monomer positions have been shifted and
rescaled to the segment [0,1]. The inset shows the same quantities
on the logarithmic scale; the dashed has slope 1.25.}
\label{fig:trans}
\end{figure}
%%%%%%%%%%%%%%%%%%%%%%%%

In Fig.~\ref{fig:trans} the results of Ref.~\cite{chat} are compared with simulations
of our model with $z=1.25$ to reproduce the exponent $\alpha=0.8$.
For better comparison the allowed interval in both cases is shifted and rescaled
to the range between 0 and 1. While the curves are quite similar, they do not coincide.
The translocation data is represented by a narrower bell-shaped  curve, and
close to the boundaries is better described by a power law with exponent
$\phi\approx 1.44$~\cite{chat}, while the curve obtained in our
simulations produces a lower exponent of $\phi\approx 1.2$.  The differences
between the two behaviors is better observed on the logarithmic scale
in the inset of Fig.~\ref{fig:trans}. Thus, there are quantitative differences
between translocation and anomalous diffusion of a monomer
with a similar exponent $\alpha$.

%%%%%%%%%%%%%%%%%%%%%%%%%%%%%%%%%%%%%%%%%%%%%%%%%%%%%%%%%%%%%%%%%%%%%

\subsection{Summary}\label{sec:disc}

In this work we concentrated on a group of subdiffusion processes in which
a tunable anomalous exponent $\alpha$ is generated through collective
behavior of many degrees of freedom. This is achieved by superposition of
linear modes in which the relaxation times are scaled by a power law.
In the polymer language this corresponds to following a tagged monomer
when the friction coefficients of the Rouse modes have a power-law
dependence on wavelength. In the absence of absorbing boundaries the
model can be solved exactly; starting from a point the PDF of the anomalous
walker is a Gaussian whose width grows in time as $t^{\alpha}$. We were not
able to solve the problem in the presence of absorbing boundaries,
and resorted to numerical simulations.
With a single absorbing point the PDF of absorption decays slowly at
long times as $t^{-1-\theta}$. The power law decay can be justified by noting
that the particle can avoid absorption by moving away from the trapping point.
A qualitative understanding of the behavior is not yet attained: Estimates
based on the fractional Fokker-Planck equation suggest \cite{brd}
$\theta=\alpha/2$, while an alternative picture from fractional Brownian
motion suggests \cite{krug} $\theta=1-\alpha/2$. Reference \cite{krug}
provides provides numerical support for the latter, while our results
are more consistent with the former. The possibility of systematic errors
prevents us from making a definite statement on this point, and indicate
necessity of further work. When the tagged monomer is confined to an interval
bounded by two traps, the survival probability is found to decay exponentially
at long times, irrespective of the subdiffusive exponent.

An interesting feature of the process is the vanishing of the PDF on
approaching an absorbing boundary (whether single or double). The method
of images, which is only valid for regular (Markovian) diffusion, predicts
a linear approach to zero, while our simulations indicate a singular form
characterized by an exponent $\phi$ for anomalous walks. While we cannot
determine this exponent precisely due to various sampling problems, our
results do not appear to support recently proposed exponent relations
\cite{zrm}. The similarities in the shapes of the stable PDFs of surviving
walkers in an interval for different values of $\alpha$, initially raised
the hope that they can be collapsed by a simple transformation (e.g.\
raising them to some power). However, the mismatch between the curves
obtained for different exponents $\alpha$ is sufficient to rule out an
over-arching super-universality. Furthermore, the discrepancies between
translocation of a self-avoiding polymer and an anomalous diffuser with
a similar exponent, suggest that the exponent $\alpha$ is not sufficient to
characterize universality. The situation is reminiscent of critical
phenomena in which Gaussian (linear) models can be devised to reproduce a
particular critical exponent, but which do not capture the full complexity of
the non-linear theory. The absence of definitive agreement between numerics and proposed models is on one hand disappointing, but on the other hand points to the necessity of further work and clarification.

The linear nature of the underlying model raises the hope that exact
solutions may be within reach. In the meantime the model does provide
a means of generating anomalous walkers with a tunable exponent that
incorporate some realistic features of collective dynamics of interacting
degrees of freedom. There are certainly  puzzles pertaining to the
behavior of such anomalous walkers close to an absorbing boundary. To
answer these questions simulations need to probe sufficiently short
times to remain in the regime of anomalous dynamics, but long enough
to ensure convergence to stable forms. Our simulations had a rather
limited range of times satisfying the above constraints. While one
order of magnitude increase in $N$ could open a broad range of
validity of the above conditions, it would significantly slow down the
simulations. In addition, working at longer times significantly
increases the attrition of the samples, and requires increasing sample
size by several orders of magnitude. Currently, such ideal conditions
are beyond our numerical abilities, but some improvement over the
current results are certainly possible.

\begin{acknowledgments}
We thank S. M. Majumdar, A. Rosso, and A. Zoia for useful
advice and discussions. This work was supported by
the Israel Science Foundation Grant No. 99/08 (YK) and
by the National Science Foundation Grant
No. DMR-08-03315 (MK). Part of this work was carried out at the Kavli
Institute for Theoretical Physics, with support from NSF Grant
No. PHY05-51164 (MK \& YK).
 \end{acknowledgments}

\end{document}